\documentclass[twocolumn,floatfix,prl]{revtex4}%
\usepackage{graphicx}%
\usepackage{amsmath}%
\usepackage{amsfonts}
\usepackage{amssymb}
\usepackage{mathrsfs}
\usepackage{color}
\usepackage{soul}
\usepackage{booktabs}
\usepackage{bm}
\def\a{{ {\alpha} }}
\def\k{{ {\bm k} }}

\def\q{{ {\bm q} }}

\def\0{{ {\bm 0} }}
\def\v{{ {\bm v} }}
\def\r{{ {\bm r} }}
\def\R{{ {\bm R} }}

\allowdisplaybreaks[4]

\begin{document} 
\title{
Prominent Intrinsic Orbital Hall Effect in the 135 Kagome Metal Family and 
Orbital Responses in the Loop-Current Phase
}
\author{
Jianxin Huang$^{1*}$, Rina Tazai$^2$, Youichi Yamakawa$^1$, and Hiroshi Kontani$^1$}

\date{\today }

\begin{abstract}
The study of electronic orbital degrees of freedom, including the generation and control of orbital currents and orbital angular momentum, has emerged as a vibrant research field. 
Here, we study the orbital Hall effect (OHE), one of the key mechanisms for orbital current generation, in transition-metal kagome-lattice metals.
We propose a large positive OHE in CsTi$_3$Bi$_5$ and negative OHEs in CsV$_3$Sb$_5$ and CsCr$_3$Sb$_5$ models. 
Orbital-sector decomposition shows that the $\vert l^z_d \vert=2$ $d$-orbital channel gives a positive contribution, whereas the $\vert l^z \vert=1$ $d$- and $p$-orbital channels can give negative contributions.
Control calculations suggest a persistent orbital-sector sign tendency near the actual filling, while strong $p$-$d$ hybridization modulates the quantitative balance and total OHE sign.
Thus, the compound and filling dependence of the OHE reflects both the sign tendency of each orbital sector and the hybridization-controlled balance among them.
Furthermore, we investigate the loop-current phase of CsV$_3$Sb$_5$ and show that it induces finite local atomic orbital angular momentum.
We also show that loop-current-related symmetry lowering allows finite symmetric components of the orbital conductivity tensor.
This study provides a basis for exploring orbital currents and local orbital-angular-momentum responses in strongly correlated kagome metals.
\end{abstract}

\affiliation{
$^1$Department of Physics, Nagoya University,
Nagoya 464-8602, Japan
\\
$^2$RIKEN Center for Emergent Matter Science (CEMS), RIKEN,
Wako, Saitama 351-0198, Japan 
}
 
\sloppy

\maketitle

\subsection{I. INTRODUCTION} \vspace{-0.5em}

In recent years, the exploration of electronic orbital degrees of freedom—beyond charge and spin—has led to the rapid development of a new research field. This emerging field has witnessed significant progress on both theoretical and experimental fronts, opening new avenues for understanding and controlling quantum transport phenomena \cite{OHE-rev1, OHE-rev2}.

\begin{figure}[htp]
\includegraphics[width=.8\linewidth]{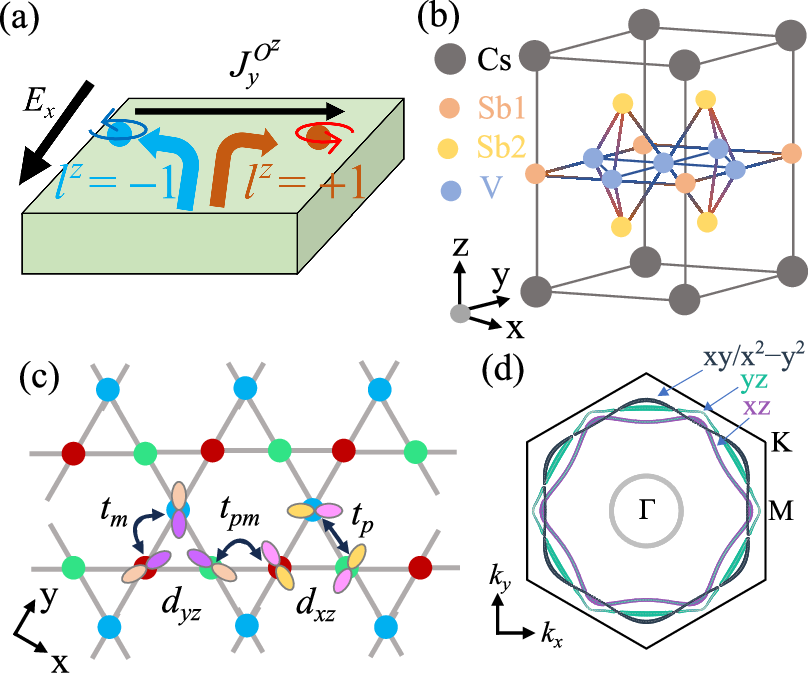}
\caption{
(a) Schematic figure of the orbital Hall effect. Under an external electric field, electrons with different orbital angular momenta move to different sides, leading to the generation of an orbital current.
(b) Crystal structure of CsV$_3$Sb$_5$. The same lattice structure applies to CsCr$_3$Sb$_5$ and CsTi$_3$Bi$_5$ with atomic replacement.
(c) $d_{xz}+d_{yz}$ orbital tight-binding model for CsV$_3$Sb$_5$.
(d) Fermi surfaces of the CsV$_3$Sb$_5$ $30$ orbital tight-binding model derived from first-principles calculations. Kagome metals possess characteristic multiple Fermi surfaces composed of $d$- and $p$-orbitals.
}

\label{fig:fig1}
\end{figure}

A central concept in this field is the orbital Hall effect (OHE)—a phenomenon in which a transverse flow of orbital angular momentum arises in response to an external electric field, in the absence of an external magnetic field, as shown in Fig. \ref{fig:fig1} (a). 
Since its initial theoretical proposals around 2005 \cite{OHE-first}, the OHE has been investigated in a variety of multi-orbital systems, notably in transition metals \cite{OHE-TM, OHE-Sr2RuO4, OHE-4d5d, OHE-Si-05, OHEvsSHE1, OHEvsSHE2, OHE-Pt, OHE-SHE-Bi, OHE-Sr2RuO4-2, OHE-IB1, OHE-TMDC-1, OHE-TMDC-2, OHE-TMDC-3, OHE-2D-insulator, OHE-chiral, OHE-Ti-exp1, OHE-Ge-exp, OHE-Cr-exp, OHE-Ti-exp2, OHE-XIV, OHE-p, OHE-FPC}.
For comparison, the spin Hall effect (SHE), analogous to the OHE but producing a spin current, has been extensively investigated \cite{SHE-rev1, SHE-rev2, SHE-FM1, SHE-FM2, SHE-Semicon, SHE-kon, SHE-intrinsic1, SHE-intrinsic2, SHE-Bi1, SHE-Bi2}. The OHE and SHE originate from the orbital and spin Berry curvatures, respectively.
However, SHE relies on spin-orbit coupling (SOC), whereas OHE can arise without it according to Refs. \cite{OHE-Si-05, OHEvsSHE1, OHEvsSHE2, OHE-4d5d, OHE-TM}. Therefore, OHE is one order of magnitude larger than the SHE in various materials. 
In fact, OHE serves as a precursor to SHE in the intrinsic mechanism, which arises concomitantly with OHE in the presence of SOC \cite{OHE-TM, OHEvsSHE1, OHEvsSHE2}. 

Experimental signatures of the OHE have been observed through orbital accumulation at surfaces using the magneto-optical Kerr effect \cite{OHE-Ti-exp1, OHE-Cr-exp}, and growing attention has been directed toward the generation of orbital torques \cite{OHE-ST-FMR, OHE-OT1, OHE-OT2, OHE-app1, OHE-app2}.
Because these responses are governed by the polarity of the orbital current, the sign of the OHE is an essential characteristic of orbital transport. It determines the polarity of the transverse orbital current, the sign of orbital accumulation at opposite sample edges, and the polarity expected in inverse-OHE \cite{IOHE-APE-exp, OHE-app1, OHE-Ge-exp} and orbital-torque measurements.
Although negative SHEs are frequently predicted, negative OHEs have rarely been encountered in previous studies.
Previous studies have predicted positive OHEs in several $d$-electron systems, such as transition metals \cite{OHE-TM, OHE-4d5d, OHEvsSHE2, OHE-FPC}, and experiments have verified the predictions \cite{OHE-Ti-exp1, OHE-Cr-exp, OHE-Ti-exp2, IOHE-APE-exp}. 
Meanwhile, for various $p$-electron systems, both positive and negative OHEs have been reported \cite{OHE-XIV, OHE-p, OHEvsSHE2}. However, in $p$-electron systems with wave functions having a larger spatial spread, the importance of the anomalous position has been emphasized in Refs. \cite{APE-Si, APE-d}.

In addition, orbital degrees of freedom have gained significance, especially in unconventional superconductivity and quantum phase transitions.
In iron-based superconductors, a nonmagnetic nematic phase with $C_4$ symmetry breaking is linked to $d$-orbital ordering \cite{Iron-C4}. 
More recently, attention has shifted to nonlocal orbital ordering phenomena, 
particularly in kagome metals $A$V$_3$Sb$_5$ ($A$ = Cs, Rb, K), since around 2020 \cite{kagome-exp1, kagome-exp2, NMR1, STM1, STM2, Pressure-dep, Tazai-DW, Mix-nem, CDW-nem, kagome-Z3, LC-Mag-AHE, LC-Trans, Mag-uSR, TRSB-uSR, LC-uSR, LC-STM, Uni-Mag, MOKE, TRSB-Kerr, CDW-AHE, K-AHE, kagome-vHS}.
In kagome metals, geometrical frustration and strong electron correlations generate unconventional quantum states, such as orbital-selective bond order \cite{kagome-exp1, kagome-exp2, NMR1, STM1, STM2, Pressure-dep, Tazai-DW, Mix-nem} and orbital loop currents (LC) \cite{LC-first, kagome-Z3, CDW-nem}. 
Various theoretical studies have elucidated this mechanism and explained several electronic and transport phenomena \cite{Tazai-DW, kagome-Z3, LC-Mag-AHE, LC-Trans, Mix-nem}.
In particular, LC breaks time-reversal symmetry and induces both local and uniform magnetization \cite{LC-Mag-AHE, LC-Trans} observed by various experiments \cite{LC-uSR, TRSB-uSR, Mag-uSR, LC-STM, Uni-Mag, MOKE, TRSB-Kerr}, including the anomalous Hall effect (AHE) \cite{CDW-AHE,K-AHE}.

In this article, we investigate the OHE in the $135$ kagome-metal family using $30$-orbital tight-binding models derived from first-principles calculations.
Kagome metals provide a rich multi-orbital platform for orbital transport, yet their intrinsic OHE has not been systematically explored.
Using the Kubo formula without the anomalous-position correction \cite{OHE-formula, OHE-tor-formula, OHE-two-terms, OHE-derive, AHE-kon, AHE-rev, APE-d}, we obtain rare negative OHEs in CsV$_3$Sb$_5$ and CsCr$_3$Sb$_5$, and a large positive OHE in CsTi$_3$Bi$_5$.
Orbital-sector decomposition shows that both $p$ and $d$ orbitals contribute substantially: the $\vert l^z_d \vert=2$ $d$-orbital channel gives a positive contribution, whereas the $\vert l^z \vert=1$ $d$- and $p$-orbital channels can give negative contributions.
Thus, the compound and filling dependence of the OHE reflects the balance among orbital-sector contributions.
Momentum-space orbital Berry-curvature analysis points to $p$-$d$-hybridized band regions, and a further control calculation for CsV$_3$Sb$_5$ suggests that strong $p$-$d$ hybridization quantitatively modulates this balance and plays an important role in controlling the material-dependent OHE sign.

Beyond the OHE, we also demonstrate that the LC state induces local atomic orbital angular momentum on the $d$ atoms.
Although the $d$-atomic orbital angular momentum transferred from the LC is small, it may provide a new principle for strongly correlated orbitronics.
We also show that LC-related symmetry lowering allows finite symmetric components of the orbital conductivity tensor.

\vspace{-0.8cm}

\subsection{II. MODEL HAMILTONIAN} \vspace{-0.5em}

\begin{figure}[htp]
\includegraphics[width=.8\linewidth]{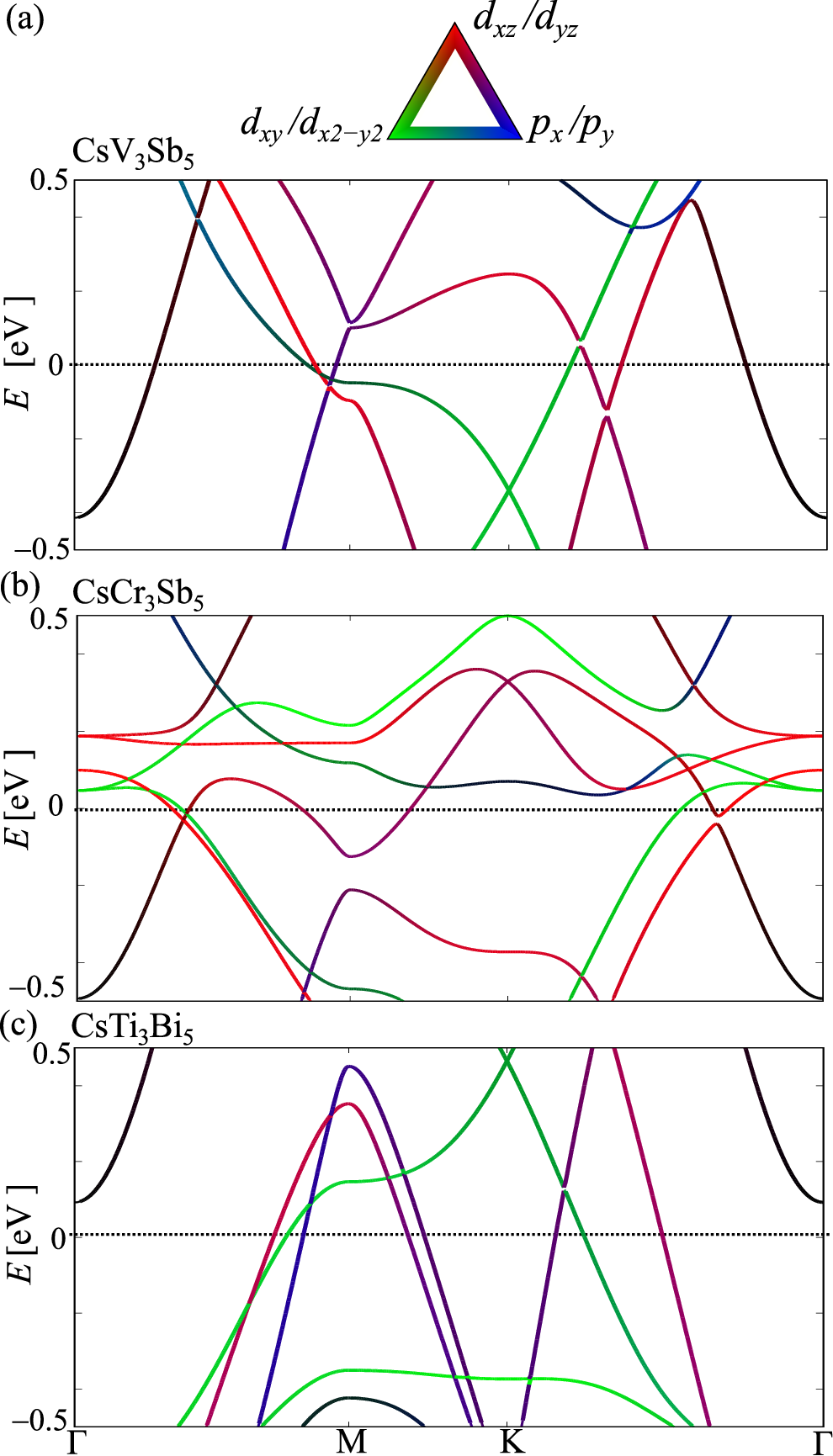}
\caption{
(a) (b) (c): Band structure of (a) CsV$_3$Sb$_5$ (n=31), (b) CsCr$_3$Sb$_5$ (n=34) and (c) CsTi$_3$Bi$_5$ (n=28). CsV$_3$Sb$_5$ presents the van Hove singularity near the Fermi level, while in CsTi$_3$Bi$_5$ the vHS is far from the Fermi level. CsCr$_3$Sb$_5$ exhibits flat bands.
The orbital characters are represented by the triangular guide above the panels, whose vertices correspond to $d_{xz}/d_{yz}$, $d_{xy}/d_{x^2-y^2}$, and $p_{x}/p_{y}$ components.
}
\label{fig:band}
\end{figure}

The three-dimensional (3D) kagome lattice structure of the $135$ kagome-metal family is shown in Fig. \ref{fig:fig1} (b). Taking CsV$_3$Sb$_5$ as an example, each unit cell is composed of three V-ion sublattices and five Sb-ion sublattices. 
The $135$ kagome-metal family studied here includes the well-known kagome metal CsV$_3$Sb$_5$, as well as the recently found CsTi$_3$Bi$_5$ and CsCr$_3$Sb$_5$. 
We derive the $30$-orbital tight-binding model with $15$ transition-metal $d$ orbitals and $15$ pnictogen $p$ orbitals for V-Sb, Cr-Sb, and Ti-Bi compounds based on the band structure obtained using WIEN$2$K software. Spin-orbit coupling is neglected in this work. To confirm the reliability of the tight-binding models, we compare the original first-principles band structures with the corresponding three-dimensional tight-binding models in Appendix A. Due to the small interlayer hopping integrals, we consider only the $2$D model. 

Figure \ref{fig:fig1}(c) illustrates the kagome lattice with V-site $d_{xz}$ and $d_{yz}$ orbitals, which exhibit van Hove singularities (vHSs) near the Fermi level \cite{kagome-vHS} and play a significant role in the orbital magnetization $M_{\rm orb}$ \cite{LC-Mag-AHE, LC-Trans}. The bond order (BO) and the LC order occur primarily in the $d_{xz}$ orbitals \cite{Tazai-DW, kagome-Z3, Mix-nem}.
Figure \ref{fig:fig1}(d) shows the Fermi surfaces of the CsV$_3$Sb$_5$ tight-binding model.
These orbitals give rise to angular momentum through the linear combination
$\vert l^z_d=\pm1 \rangle\propto\vert d_{xz}\rangle \pm i \vert d_{yz}\rangle$,
$\vert l^z_d=\pm2 \rangle\propto\vert d_{x^2-y^2}\rangle \pm i \vert d_{xy}\rangle$, and
$\vert l^z_p=\pm1 \rangle\propto\vert p_x\rangle \pm i \vert p_y\rangle$.

In the present analysis, the kagome lattice mainly provides a multi-sublattice, multi-orbital electronic structure with characteristic vHSs, weakly dispersive bands, band crossings, and strong $d$-$p$ hybridization near the Fermi level.
These features enhance and redistribute the orbital Berry curvature, particularly through the $p_{x/y}$-$d_{xz/yz}$ hybridization discussed below.
Thus, the $135$ compounds are best viewed as kagome-derived multi-orbital platforms for orbital transport, rather than as realizing a kagome-exclusive OHE mechanism.

Changing the transition-metal element changes the electron filling: $n=28$ for CsTi$_3$Bi$_5$, $n=31$ for CsV$_3$Sb$_5$, and $n=34$ for CsCr$_3$Sb$_5$.
Figure \ref{fig:band} shows the corresponding tight-binding band structures, where the colors indicate the orbital weights.
The purple regions indicate sizable hybridization between the $p_{x/y}$ and $d_{xz/yz}$ orbitals.
In CsV$_3$Sb$_5$, a van Hove singularity (vHS) lies near the Fermi level (Fig. \ref{fig:band}(a)).
In CsCr$_3$Sb$_5$, the additional electron per Cr atom shifts the Fermi level upward by about $0.5$ eV and places it near flat bands (Fig. \ref{fig:band}(b)).
In CsTi$_3$Bi$_5$, the Fermi level is lowered by about $0.8$ eV, and the vHS is displaced from the Fermi energy (Fig. \ref{fig:band}(c)).

In addition to the transition-metal element, the ligand species also differs among the models: the V-Sb and Cr-Sb models involve Sb $5p$ orbitals, whereas the Ti-Bi model involves Bi $6p$ orbitals.
A previous study of Bi$_{1-x}$Sb$_x$ revealed a dramatic change in SHE with Sb concentration $x$ \cite{SHE-Bi1, SHE-Bi2}, implying a significant difference in the $p$ orbital between Bi and Sb ions.
In the Sec. IV, we discuss that this difference may contribute to the variation in OHE.

Although their band structures differ, these kagome metals exhibit intriguing quantum phases, such as the BO and LC states in CsV$_3$Sb$_5$ \cite{kagome-Z3, LC-Mag-AHE}, the recently predicted odd-parity BO in CsTi$_3$Bi$_5$, which induces the nonlinear Hall effect \cite{Ti-BO}, and a novel stripe CDW order in CsCr$_3$Sb$_5$ \cite{Cr-CDW}.
These quantum states provide fascinating platforms for studying orbital degrees of freedom.
In the latter part of this work, we focus on the orbital angular momentum and orbital transport associated with the BO and LC states in CsV$_3$Sb$_5$, introduced through real and imaginary hopping modulations, respectively, in Secs. V and VI.

\subsection{III. FORMULATIONS} \vspace{-0.5em}

In linear response, the current and orbital flow are written as follows:
\begin{eqnarray}
    J_{\mu}=\sigma_{\mu\nu}E_{\nu} \\
    J^{O^{\alpha}}_{\mu}=O^{\alpha}_{\mu\nu}E_{\nu}
\end{eqnarray}
where $\alpha, \mu, \nu=x, y, z$, $E_{\nu}$ is the external electric field, $J_{\mu}$ is the current density, and $J^{O^z}_{\mu}$ is the current density of the orbital angular momentum. Here, $\a$ indicates the direction of the orbital angular momentum, and we set $\alpha=z$ hereafter. 
The current operator is given by $\hat{J}_{\mu}=(-e)\hat{v}_{\mu}$ and the orbital angular momentum current density operator is expressed as $\hat{J}^{O^z}_{\mu}=(\hat{v}_{\mu}l^z+l^z\hat{v}_{\mu})/2$ \cite{OHE-TM}. Here, $\hat{v}_{\mu}=\partial_{k_{\mu}} \hat{H}$ is the velocity operator and $l^z$ is the orbital angular momentum matrix. The unit of energy is eV, and we set $\hbar=k_{\rm B}=1$ unless otherwise stated.

Previous studies derived the formula for intrinsic anomalous Hall conductivity, including dissipation effects based on the Kubo formalism \cite{OHE-formula}, which can be extended to OHE by replacing the corresponding operator. 
The conductivity $\sigma_{xx}$ and $O^z_{xy}$ can be obtained from the current-current correlation function $\langle J_x(0)J_x(t)\rangle$ and the orbital current-current correlation function $\langle J^{O^z}_x(0)J_y(t)\rangle$, respectively, where $t$ denotes the time. 

We denote the symmetric and antisymmetric components of the conductivity tensor by superscripts $+$ and $-$, respectively:
\begin{eqnarray}
    \sigma^{\pm}_{\mu\nu}=(\sigma_{\mu\nu}\pm\sigma_{\nu\mu})/2, \\
    O^{\pm}_{\mu\nu}=(O_{\mu\nu}\pm O_{\nu\mu})/2
\end{eqnarray}
For antisymmetric components, the relation $\sigma^-_{\mu\mu}=O^-_{\mu\mu}=0$ holds, and $O^-_{xy}=-O^-_{yx}$ is referred to as the intrinsic orbital Hall conductivity. Similarly, the symmetric part has the relation $O^+_{xy}=O^+_{yx}$.

In this work, a band-independent relaxation time $\tau$ is introduced for the conduction electrons. The corresponding band independent quasiparticle damping rate is given by $\gamma\equiv\hbar/2\tau$. We assume that $\tau$ is sufficiently large and calculate the leading divergent term in the transport coefficient with respect to $\tau$.

In the low-temperature limit and for small $\gamma$, the leading divergent term of the antisymmetric component of OHE is given by \cite{OHE-formula, OHE-tor-formula, OHE-two-terms, OHE-derive, SOT-rev, tensor-symmetry}:
\begin{equation}
    O^{-}_{xy}=-\frac{2}{N_{\k}}\sum_{\k}\sum^{\rm occ}_a\sum_{b\ne a}
    \frac{{\rm{Im}}\{J^{O^z}_{\k,ab,x}v_{\k,ba,y}\}}
    {(E_{a,\k}-E_{b,\k})^2}.
    \label{eqn:T-even}
\end{equation}
where $a$ and $b$ are band indices, and $E_{a(b),\k}$ and $v_{\k,ba,y}$ are the corresponding band energy and velocity matrix element, respectively. $N_k$ is the number of $\k$ points used in the Brillouin-zone summation, which is set to $N_k=600\times600$ in this work.
Note that $O^{-}_{\mu\nu}$ reflects the Berry-phase-related structure of the Bloch states, corresponding to the intrinsic Fermi-sea contribution \cite{AHE-rev}.

On the other hand, the symmetric component is expressed as follows:
\begin{equation}
    O^{+}_{\mu\nu}=\frac{-1}{\pi N_{\k}}\sum_{\k,a,b}
    \frac{{\rm{Re}}\{\gamma^2 J^{O^z}_{\k,ab,\mu}v_{\k,ba,\nu}\}}
    {(E_{a,\k}^2+\gamma^2)(E_{b,\k}^2+\gamma^2)}.
    \label{eqn:T-odd}
\end{equation}
The symmetric contribution is proportional to $\tau^1$ in a system without band degeneracy: when $a=b$, the energies in the formula can become $0$ at the same time, and the behavior $1/\gamma$ ($\tau^1$) occurs after the $k$-summation. This part represents a scattering-dependent Fermi-surface contribution.

Following the conventional terminology used for the SHE in isotropic models \cite{SHE-rev1}, we refer to the antisymmetric transverse component $O^-_{xy}$ as the OHE in the high-symmetry kagome phase and focus on this contribution in Sec. IV. In lower-symmetry phases, symmetric orbital conductivity components can also be allowed, representing additional orbital-current responses beyond the conventional antisymmetric OHE. The loop-current bond-ordered phase studied here provides such an example, and the symmetric component $O^+_{xy}$ is discussed separately in Sec. VI.
The details of the derivation for the two components are explained in Appendix B. 

For arbitrary $\gamma$, the Kubo formula can be decomposed into one Fermi-surface term $O^I_{xy}$ and two Fermi-sea terms $O^{IIa}_{xy}$ and $O^{IIb}_{xy}$ \cite{OHE-4d5d}:
\begin{align}
    &O^I_{xy}=\frac{-1}{2\pi N_k}\sum_{\k,a \ne b} \{J^{O^z}_{\k,ab,x}J_{\k,ba,y}\} \notag \\
    &\hspace{4em}\times\left[ \frac{1}{(E_{b,\k}-i\gamma)(E_{a,\k}+i\gamma)} \right], \label{eqn:O^I} \\
    &O^{IIa}_{xy}=\frac{-1}{2\pi N_k}\sum_{\k,a \ne b} {\rm Im}\{J^{O^z}_{\k,ab,x}J_{\k,ba,y}\}\frac{1}{E_{b,\k}-E_{a,\k}} \notag \\
    &\hspace{4em}\times {\rm Im}\left[ \frac{E_{b,\k}+E_{a,\k}-2i\gamma}{(E_{b,\k}-i\gamma)(E_{a,\k}-i\gamma)} \right], \label{eqn:O^IIa} \\
    &O^{IIb}_{xy}=\frac{1}{\pi N_k}\sum_{\k,a \ne b} {\rm Im}\{J^{O^z}_{\k,ab,x}J_{\k,ba,y}\}\frac{1}{(E_{b,\k}-E_{a,\k})^2} \notag \\
    &\hspace{4em}\times{\rm Im}\left[ {\rm{ln}} \left(\frac{E_{b,\k}-i\gamma}{E_{a,\k}-i\gamma} \right) \right]. \label{eqn:O^IIb} 
\end{align}

When $\gamma\to0$, $O^I_{xy}+O^{IIa}_{xy}=0$, and $O^{IIb}_{xy}$ in Eq. (\ref{eqn:O^IIb}) reduces to the Berry-curvature contribution \cite{AHE-rev, kon-18, IIb-term}:
\begin{equation}
    O^{IIb}_{xy}=\frac1{N_k}\sum_{\k,a}f_{T=0}(E_{a,\k})\Omega^a(\k) \label{eqn:O^BC},
\end{equation}
where $\Omega^a(\k)$ represents the orbital Berry curvature expressed as
\begin{equation}
    \Omega^a(\k)=\sum_{b\ne a}\frac{2\ {\rm Im}\{J^{O^z}_{\k,ab,x}J_{\k,ba,y}\}}{(E_{a,\k}-E_{b,\k})^2}. \label{eqn:BC}
\end{equation}

However, this relation breaks for finite $\gamma$, and $O^I_{xy}\approx O^{-}_{xy}$ is valid in real metallic systems with a large damping $\gamma$ \cite{OHE-4d5d, AHE-kon, AHE-rev}. The damping-rate dependence of three terms is presented in Appendix C.

To analyze the momentum-space origin of the OHE, we use the $\k$-resolved Berry-curvature-related contribution $O^{IIb}_{xy}(\k)$, defined by $O^{IIb}_{xy}=\frac{1}{N_k}\sum_{\k}O^{IIb}_{xy}(\k)$.
In Sec. IV, we focus on this quantity as the momentum-space distribution relevant to the intrinsic OHE.


\subsection{IV. ORBITAL HALL EFFECT IN KAGOME METALS} \vspace{-0.5em}

\begin{figure}[htp]
\includegraphics[width=.8\linewidth]{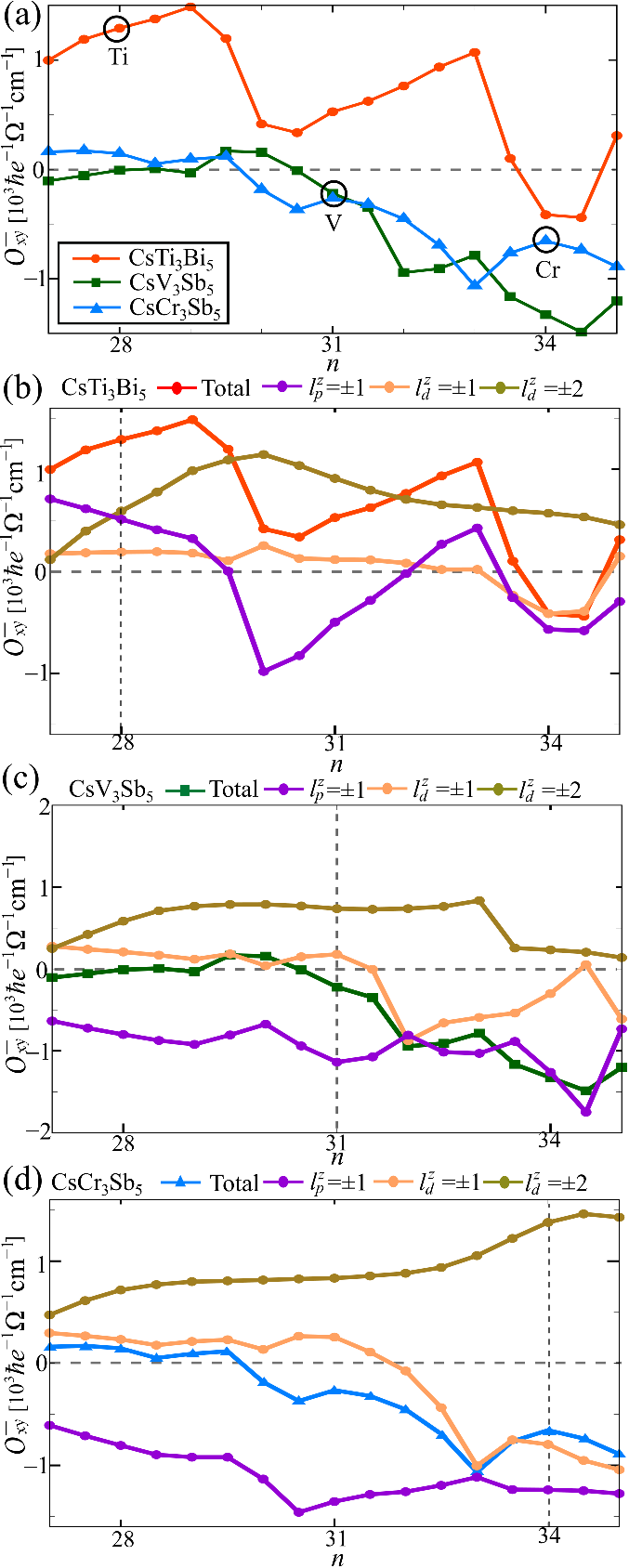}
\caption{
(a) Filling-dependent OHE $O^{-}_{xy}$ for the CsTi$_3$Bi$_5$, CsV$_3$Sb$_5$ and CsCr$_3$Sb$_5$ tight-binding models. Circles indicate the actual fillings for three materials. 
A large and positive OHE in CsTi$_3$Bi$_5$ and the negative OHEs in CsV$_3$Sb$_5$ and CsCr$_3$Sb$_5$ are predicted.
(b) (c) (d): Total OHE and its contributions from different orbital-angular-momentum channels ($\vert l^z_p \vert =1, \vert l^z_d \vert =1$ and $\vert l^z_d \vert =2$) in (b) CsTi$_3$Bi$_5$, (c) CsV$_3$Sb$_5$ and (d) CsCr$_3$Sb$_5$. The vertical dashed line marks the actual filling for each material.
}
\label{fig:OHE}
\end{figure}

We first calculate the intrinsic antisymmetric orbital Hall coefficient $O^{-}_{xy}$ for three kagome-metal models. 
The temperature is set to $T=0.001$ eV for the low-temperature approximation.

Calculations are performed by varying the electron density $n$.
Figure \ref{fig:OHE}(a) shows the filling dependence, with black circles marking the actual fillings.
At the actual filling, CsTi$_3$Bi$_5$ exhibits a large positive OHE of about $1.3\times10^3\ [\hbar e^{-1}\Omega^{-1}{\rm cm}^{-1}]$, whereas CsV$_3$Sb$_5$ and CsCr$_3$Sb$_5$ show negative OHEs of about $-0.2\times10^3$ and $-0.6\times10^3\ [\hbar e^{-1}\Omega^{-1}{\rm cm}^{-1}]$, respectively.
The magnitude can become much larger away from the actual filling, suggesting that carrier doping may be used to enhance the OHE response.
The negative signs in the Sb-based compounds indicate an orbital-current polarity opposite to that in CsTi$_3$Bi$_5$; consequently, the sign of the transverse $L^z$ accumulation and the polarity of an inverse-OHE signal are reversed for the same applied electric field \cite{OHE-app1}.

To investigate the origin of this sign difference, we decompose $O^{-}_{xy}$ into contributions from different $l^z$ matrix elements. 
Since the $l^z$ matrix has finite elements only between specific orbital pairs \cite{OHE-TM, OHE-XIV}, the OHE can be separated into three channels: $\vert l^z_d \vert =1$, $\vert l^z_d \vert =2$, and $\vert l^z_p \vert =1$. 
Their filling dependencies are shown in Figs. \ref{fig:OHE}(b)-(d).

For CsTi$_3$Bi$_5$, the $\vert l^z_d \vert =2$ channel is positive over the relevant filling range and dominates the positive OHE at $n=28$ (Fig. \ref{fig:OHE}(b)). 
The $\vert l^z_p \vert =1$ channel gives the second-largest positive contribution, whereas the $\vert l^z_d \vert =1$ channel becomes negative only at higher electron densities. 
Consequently, the total OHE remains positive at the actual filling.

For CsV$_3$Sb$_5$ and CsCr$_3$Sb$_5$, the $d$-orbital contributions are qualitatively similar to those in CsTi$_3$Bi$_5$. 
The crucial difference is the $\vert l^z_p \vert =1$ channel: in both Sb-based models, this contribution remains negative over the considered filling range and, together with the $\vert l^z_d \vert =1$ channel, overcomes the positive $\vert l^z_d \vert =2$ contribution near the actual filling.
Thus, the sign of the intrinsic OHE is not fixed by a single orbital channel, but results from the competition among several orbital-sector contributions.

\begin{figure}[htp]
\includegraphics[width=.8\linewidth]{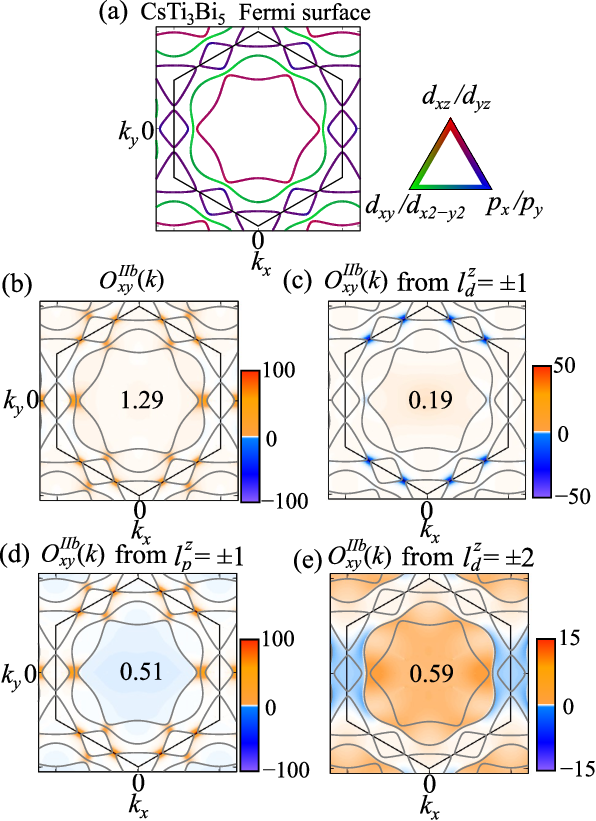}
\caption{
(a) Fermi surfaces of CsTi$_3$Bi$_5$ at $n=28$. The orbital weights are represented by the triangular guide, whose vertices correspond to $d_{xz}/d_{yz}$, $d_{xy}/d_{x2-y2}$, and $p_x/p_y$ components. (b)-(e) $\k$-space distributions of the OHE Fermi-sea term $O_{xy}^{IIb}(\k)$. The values inside the panels indicate the Brillouin-zone integrals, $\frac1{N_k} \sum_{\k} O_{xy}^{IIb}(\k)$, in units of $10^3 [\hbar e^{-1} \Omega^{-1} \rm{cm}^{-1}]$. The thin overlaid curves represent the Fermi surfaces. Panel (b) shows the total Fermi-sea contribution $O_{xy}^{IIb}(\k)$, while panels (c), (d), and (e) show the contributions from $|l_z^d|=1$, $|l_z^p|=1$, and $|l_z^d|=2$, respectively.
}
\label{fig:k2b-Ti}
\end{figure}

\begin{figure}[htp]
\includegraphics[width=.8\linewidth]{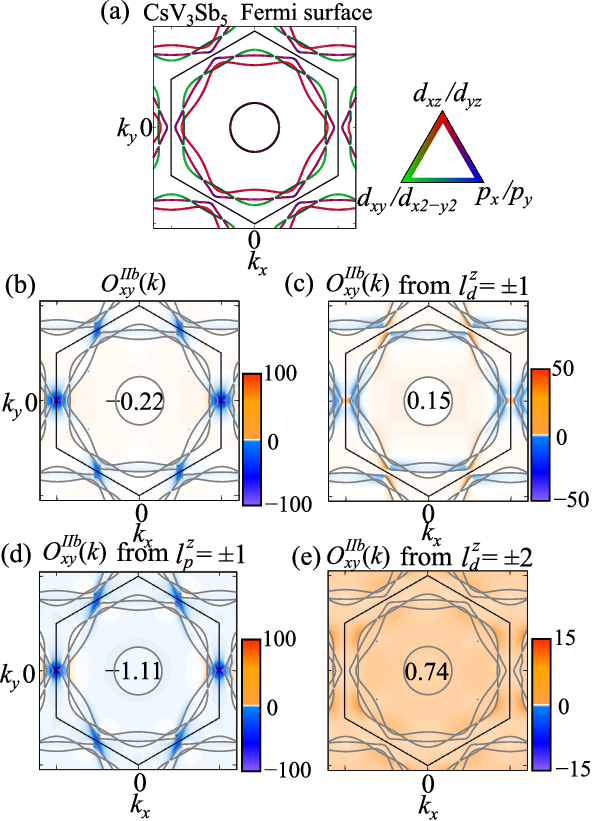}
\caption{
(a) Fermi surfaces of CsV$_3$Sb$_5$ at $n=31$. The orbital weights are represented by the triangular guide, whose vertices correspond to $d_{xz}/d_{yz}$, $d_{xy}/d_{x2-y2}$, and $p_x/p_y$ components. (b)-(e) $\k$-space distributions of the OHE Fermi-sea term $O_{xy}^{IIb}(\k)$. The values inside the panels indicate the Brillouin-zone integrals, $\frac1{N_k} \sum_{\k} O_{xy}^{IIb}(\k)$, in units of $10^3 [\hbar e^{-1} \Omega^{-1} \rm{cm}^{-1}]$. The thin overlaid curves represent the Fermi surfaces. Panel (b) shows the total Fermi-sea contribution $O_{xy}^{IIb}(\k)$, while panels (c), (d), and (e) show the contributions from $|l_z^d|=1$, $|l_z^p|=1$, and $|l_z^d|=2$, respectively.
}
\label{fig:k2b-V}
\end{figure}

\begin{figure}[htp]
\includegraphics[width=.8\linewidth]{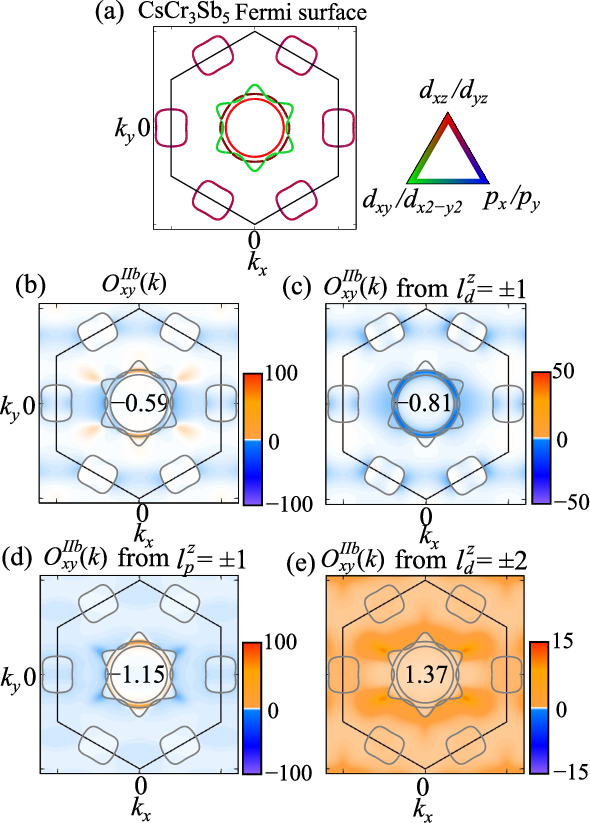}
\caption{
(a) Fermi surfaces of CsCr$_3$Sb$_5$ at $n=34$. The orbital weights are represented by the triangular guide, whose vertices correspond to $d_{xz}/d_{yz}$, $d_{xy}/d_{x2-y2}$, and $p_x/p_y$ components. (b)-(e) $\k$-space distributions of the OHE Fermi-sea term $O_{xy}^{IIb}(\k)$. The values inside the panels indicate the Brillouin-zone integrals, $\frac1{N_k} \sum_{\k} O_{xy}^{IIb}(\k)$, in units of $10^3 [\hbar e^{-1} \Omega^{-1} \rm{cm}^{-1}]$. The thin overlaid curves represent the Fermi surfaces. Panel (b) shows the total Fermi-sea contribution $O_{xy}^{IIb}(\k)$, while panels (c), (d), and (e) show the contributions from $|l_z^d|=1$, $|l_z^p|=1$, and $|l_z^d|=2$, respectively.
}
\label{fig:k2b-Cr}
\end{figure}

We next analyze the momentum-space distribution of the orbital Berry curvature to clarify where these orbital-sector contributions originate. 
As explained in Eq. (\ref{eqn:O^BC}), the Fermi-sea term $O^{IIb}_{xy}$ corresponds to the orbital Berry-curvature contribution in the limit $\gamma\to0$. 
Here, we use $\gamma=0.005$ eV and examine $O^{IIb}_{xy}(\k)$.

Figure \ref{fig:k2b-Ti} shows the Fermi surfaces with colors indicating orbital character and the $k$-space distribution of $O^{IIb}_{xy}(\k)$ for CsTi$_3$Bi$_5$ at $n=28$. The values in the panels indicate the Brillouin-zone integral $\frac1{N_k}\sum_{\k}O^{IIb}_{xy}(\k)$, and the gray lines show the Fermi surfaces.
The $O_{xy}^{\rm IIb}(\bm{k})$ distributions from the $\vert l^z_d \vert=1$ and $\vert l^z_p \vert=1$ orbitals show spot-like structures near nearly degenerate bands. 
In particular, prominent spots from both channels appear on the Brillouin-zone (BZ) boundary, reflecting strong $p$-$d$ hybridization in the $|l^z|=1$ sector. 
By contrast, the $\vert l^z_d \vert =2$ channel in Fig. \ref{fig:k2b-Ti}(e) is more broadly distributed and lacks such pronounced spot-like features. 
Nevertheless, the $k$-integrated $\vert l^z_p \vert =1$ and $\vert l^z_d \vert =2$ contributions are comparable in magnitude.

Similar $k$-space structures for the CsV$_3$Sb$_5$ and CsCr$_3$Sb$_5$ models are shown in Figs. \ref{fig:k2b-V} and \ref{fig:k2b-Cr}. 
One important difference is that the $\vert l^z_p \vert =1$ contribution changes sign in the Sb-based models (Figs. \ref{fig:k2b-V}(d) and \ref{fig:k2b-Cr}(d)). 
This sign change gives negative $k$-integrated values for the $p$-orbital channel and leads to the negative total $O^{IIb}_{xy}$ shown in Figs. \ref{fig:k2b-V}(b) and \ref{fig:k2b-Cr}(b), consistent with Figs. \ref{fig:OHE}(c) and \ref{fig:OHE}(d).

The momentum-space distributions suggest that the sign difference may be related to the $|l^z|=1$ sector involving $p_{x/y}$ and $d_{xz/yz}$ orbitals. We therefore compare material-dependent parameters relevant to this hybridization, the hopping integrals and onsite-energy separations.
Table \ref{tab:E+t} lists representative first-principles tight-binding parameters: the hopping integrals between the transition-metal-site $d_{xz}/d_{yz}$ orbitals and the nearest-neighbor ligand-site Sb2 $p_x/p_y$ orbitals, together with the corresponding onsite-energy differences ($E_{dxz}-E_{px/py}$ and $E_{dyz}-E_{px/py}$).

\begin{table}[h]
    \centering
    \caption{
    The values of the nearest-neighbor hopping integrals and energy differences between the transition-metal-site $d_{xz}/d_{yz}$ orbitals and ligand-site (Sb2 or Bi2) $p_x/p_y$ orbitals for three kagome metal TB models. For each $d$ orbital, only the largest hopping integral among its couplings to nearest neighbor $p_x$ and $p_y$ is shown. The energy differences are defined using the onsite energies, with $E_{px}=E_{py}$. The unit is eV.
    }
    \begin{tabular*}{\columnwidth}{@{\extracolsep{\fill}} lccr}
    \hline
    \hline
                        &  CsCr$_3$Sb$_5$ &  CsV$_3$Sb$_5$  &  CsTi$_3$Bi$_5$ \\
    \hline
    $|t_{dxz-py}|$        &  $0.51$   &  $0.53$   &  $0.50$ \\
    $|t_{dyz-px}|$        &  $0.58$   &  $0.56$   &  $0.48$ \\
    $E_{dxz}-E_{px/py}$ &  $-0.14$   &  $0.25$   &  $0.97$ \\
    $E_{dyz}-E_{px/py}$ &  $0.64$   &  $0.99$   &  $1.61$ \\ 
    \hline
    \hline
    \end{tabular*}
    \label{tab:E+t}
\end{table}

The hopping integrals between the $\vert l^z_d\vert=1$ and $\vert l^z_p\vert=1$ sectors vary only weakly among the three materials, whereas the onsite-energy separations are much larger in the Ti-Bi system.
Notably, within first-order perturbation theory, the corrected $d$-orbital wave function contains an admixed $p$-orbital component, whose weight scales as $(t_{dp}/\Delta E_{dp})^2$, where $t_{dp}$ is the $d$-$p$ hopping integral and $\Delta E_{dp}$ is the onsite-energy separation between the $d$ and $p$ orbitals. 
Thus, a larger $\Delta E_{dp}$ suppresses the $p$-orbital admixture into the $d$-derived states. 
This suggests that the $\vert l^z_d\vert=1$-$\vert l^z_p\vert=1$ hybridization is weaker in the Ti-Bi system than in the Sb-based systems.

The connection between the $p$-$d$ hybridization and the Berry-curvature distribution is also seen in the momentum-space maps. 
The high-intensity regions of $O_{xy}^{IIb}(\bm{k})$ from the $|l^z_p|=1$ and $|l^z_d|=1$ channels appear near similar regions on the BZ boundary, where near-degenerate bands with mixed $p_x/p_y$ and $d_{xz}/d_{yz}$ orbital character are present (Figs. \ref{fig:k2b-Ti}-\ref{fig:k2b-Cr}). 
The opposite signs of these two channel contributions in the Ti-Bi model suggest that the material-dependent $p$-$d$ hybridization may be related to the sign of the $|l^z|=1$ Berry-curvature contribution.

\begin{figure}[htp]
\includegraphics[width=.8\linewidth]{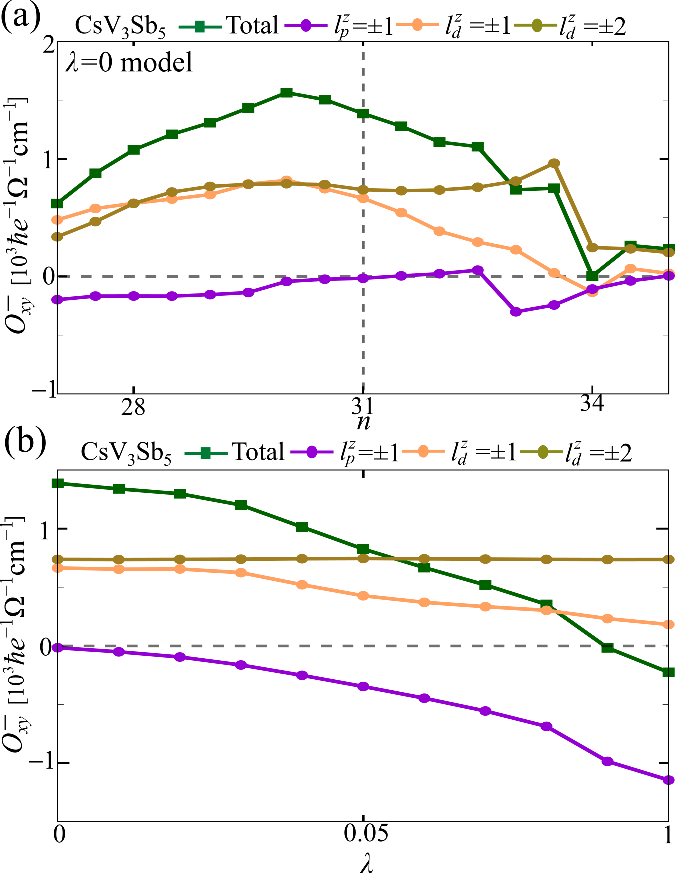}
\caption{
OHE in the CsV$_3$Sb$_5$ model with scaled hopping between Sb2 $p_{x/y}$ orbitals and V $d_{xz/yz}$ orbitals, $t^{\lambda}_{p_{x/y}-d_{xz/yz}}=\lambda t_{p_{x/y}-d_{xz/yz}}$, where $t$ is the original tight-binding hopping integral.
(a) Filling dependence of the total OHE and its orbital-sector contributions ($\vert l^z_p \vert =1$, $\vert l^z_d \vert =1$, and $\vert l^z_d \vert =2$) for $\lambda=0$, where the $p_{x/y}$-$d_{xz/yz}$ hybridization channel is switched off.
The vertical dashed line marks the actual filling $n=31$.
Around this filling, the $d$-orbital contribution is positive whereas the $p$-orbital contribution is negative, as in the original $\lambda=1$ model.
(b) $\lambda$ dependence of the total OHE and its orbital-sector contributions at $n=31$. As $\lambda$ increases, the negative $p$-orbital contribution is enhanced and eventually drives the total OHE negative.
}
\label{fig:d-p=0}
\end{figure}

To further examine the orbital-sector contributions and the role of $p$-$d$ hybridization, we perform a control calculation for CsV$_3$Sb$_5$ by scaling this hopping channel between the Sb2 $p_{x/y}$ orbitals and the V $d_{xz/yz}$ orbitals as $t^{\lambda}_{p_{x/y}-d_{xz/yz}}=\lambda t_{p_{x/y}-d_{xz/yz}}$, where $t$ is the original tight-binding hopping integral.
Figure \ref{fig:d-p=0} (a) shows the filling dependence of the OHE in the $\lambda=0$ model of CsV$_3$Sb$_5$.
Comparison with the original result in Fig. \ref{fig:OHE} shows that, around the actual filling $n=31$, the qualitative behavior of the orbital-sector contributions is unchanged: the $d$-orbital contribution is positive, whereas the $p$-orbital contribution is negative.
Thus, this sign tendency persists after switching off the $p_{x/y}$-$d_{xz/yz}$ hybridization channel, although the quantitative magnitudes and the total OHE depend on the full band structure.
This tendency is consistent with previous reports of positive OHEs in $d$-orbital systems \cite{OHE-4d5d, OHE-TM} and negative OHEs in some $p$-orbital systems \cite{OHE-XIV, OHE-p}.

Figure \ref{fig:d-p=0} (b) shows the $\lambda$ dependence of the OHE at $n=31$.
As $\lambda$ increases, stronger $p$-$d$ hybridization increases the $p$-orbital weight near the Fermi level and enhances the negative $p$-orbital contribution.
This negative contribution eventually overcomes the positive $d$-orbital contribution, leading to the negative total OHE in the original model.
Therefore, the strong $p$-$d$ hybridization in kagome metals quantitatively modulates the balance between positive $d$-orbital and negative $p$-orbital contributions and plays an important role in the OHE sign difference.

In summary, the orbital-sector decomposition shows that the OHE in the $135$ kagome-metal models arises from competing contributions with different signs: the $\vert l^z_d \vert=2$ $d$-orbital channel is positive, whereas the $\vert l^z \vert=1$ $d$- and $p$-orbital channels can be negative. Thus, the nontrivial compound and filling dependence of the OHE in kagome-metal models reflects changes in the balance among these orbital-sector contributions.
The momentum-space orbital Berry-curvature analysis points to the relevance of $p$-$d$-hybridized band regions.
The control calculation for CsV$_3$Sb$_5$ shows that the $d$-positive/$p$-negative sign tendency near the actual filling is preserved for $\lambda=0$, while the quantitative balance and the total OHE are strongly modulated by the $p_{x/y}$-$d_{xz/yz}$ hybridization.
Thus, strong $p$-$d$ hybridization is an important ingredient controlling the material-dependent OHE sign, although the microscopic origin of the individual orbital-sector signs remains to be clarified.

Finally, we comment on the anomalous position effect. 
In the present work, the orbital-current operator is evaluated with only the canonical position operator.
Recent studies have shown that the anomalous-position correction, a gauge correction arising from dipole matrix elements between Wannier basis functions (see Appendix D), can modify the magnitude and even the sign of the OHE, especially for spatially extended $p$ orbitals \cite{APE-d, OHE-4d5d, IOHE-APE-exp, OHE-FPC}.
Since sizable $p$-orbital contributions are involved here, the robustness of the predicted OHE values and signs should be examined in future calculations that include this correction.
Nevertheless, the present analysis identifies the intrinsic OHE, the orbital-sector decomposition, the momentum-space Berry-curvature distribution, and the importance of $|l^z|=1$ $d$-$p$ hybridization without the anomalous-position correction.


\subsection{V. $d$-ORBITAL ANGULAR MOMENTUM BY LOOP CURRENT PHASE}

\begin{figure}[htp]
\includegraphics[width=.8\linewidth]{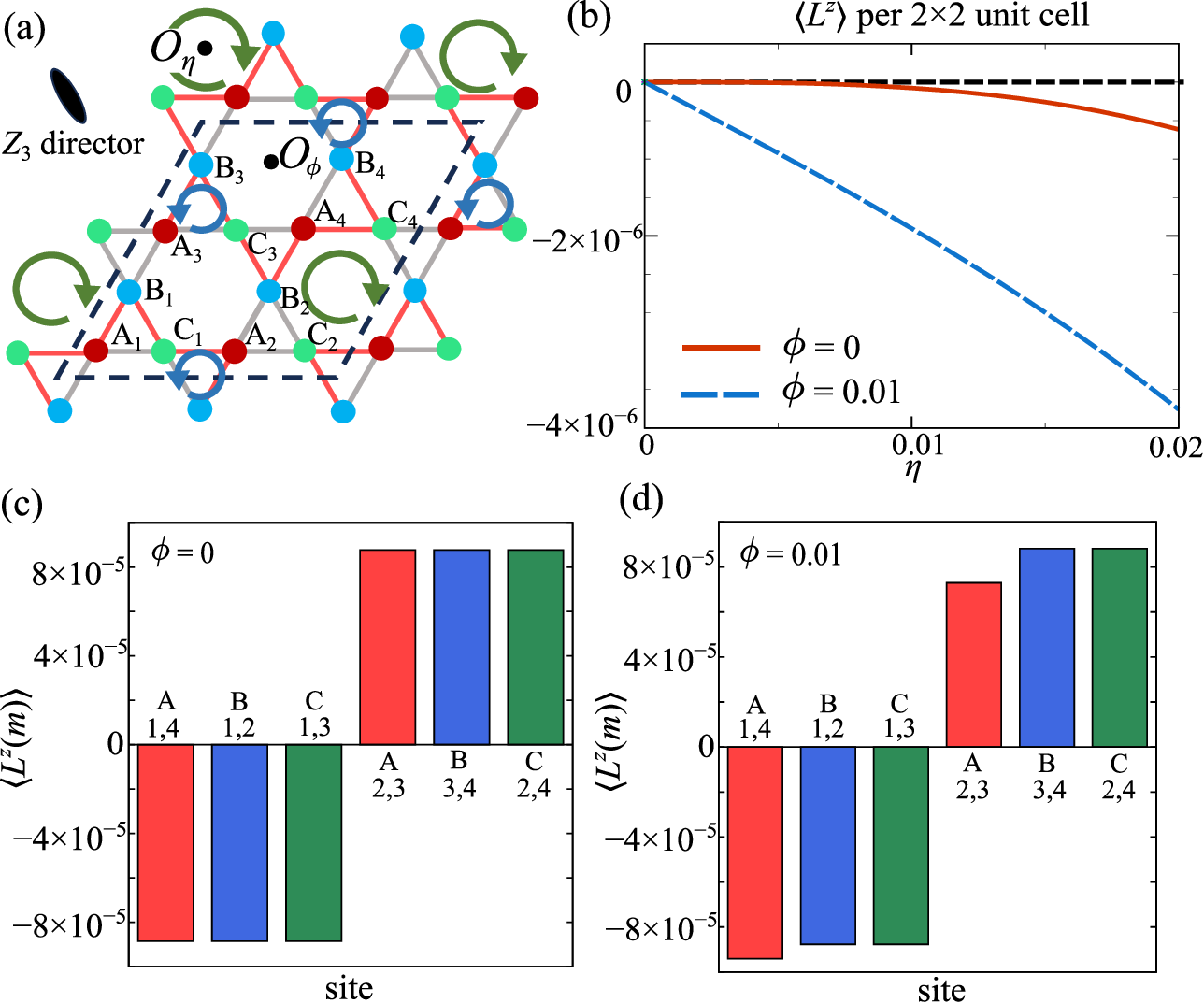}
\caption{
(a) $2\times2$ unit-cell model of CsV$_3$Sb$_5$ with LC and BO. Nematicity appears in the LC+BO coexisting state.
(b) LC-induced local atomic orbital angular momentum $\langle L^z\rangle$ per $2\times2$ unit cell as a function of $\eta$. The cases without BO ($\phi=0$) and with BO ($\phi=0.01$) are shown.
(c), (d) LC-induced site-resolved local atomic $d$-orbital angular momentum $\langle L^z(m)\rangle$, where $m$ labels the V site. The cases without BO ($\phi=0$) and with BO ($\phi=0.01$) are shown in (c) and (d), respectively.
}
\label{fig:OAM}
\end{figure}

Beyond the orbital Hall effect, the CsV$_3$Sb$_5$ tight-binding model exhibits other intriguing orbital-related properties due to the LC order. Here we examine the local atomic orbital angular momentum induced in this ordered state. In the $2\times2$ unit cell shown in Fig. \ref{fig:OAM} (a), the LC and BO are introduced by imaginary and real hopping modulations, $\delta t^c=\pm i\eta$ and $\delta t^b=\pm\phi$, respectively, in the V-site $d_{xz}$ orbital.
Here, $\eta$ ($\phi$) is the order parameter of the LC state (BO), and the two orders are explained in detail in Appendix E.

In kagome metals, conduction electrons exhibit circular motion induced by the LC order in the $d_{xz}$ orbital. From symmetry considerations, such LC order can couple to the local atomic orbital degrees of freedom.
We therefore calculate the expectation value of the total local atomic orbital angular momentum over the sites and orbitals in the enlarged $2\times2$ unit cell:
$\langle L^z\rangle=\sum_{m}\sum_{ij}\langle c_{mi}^{\dagger}(l^z)_{ij}c_{mj}\rangle$,
where $m$ labels the atomic site and $i$ and $j$ denote the local atomic orbitals included in the tight-binding model.
Here $l^z$ is the atomic orbital angular momentum matrix introduced in Sec. III.
This quantity represents the induced local atomic orbital-angular-momentum response associated with the LC state.
In this definition, $\langle L^z\rangle$ is expected to become finite in the LC state, particularly in the $|l^z_d|=1$ component associated with the $d_{xz}$ and $d_{yz}$ orbitals.

Figure \ref{fig:OAM} (b) shows the $\eta$ dependence of the obtained uniform $d+p$ orbital angular momentum $\langle L^z\rangle$ per $2\times2$ unit cell. 
Here we show two cases: one with only the LC phase ($\phi=0$) and the other with the LC$+$BO ($\phi=0.01$) coexisting state. 
For the coexisting state, $\langle L^z \rangle$ increases linearly with $\eta$. 
This result supports the expectation that $\langle L^z \rangle$ increases in proportion to the uniform orbital magnetization $M_{\rm orb}$ (defined in Refs. \cite{LC-Mag-AHE, LC-Trans}).
In fact, a previous theoretical study \cite{LC-Mag-AHE} reveals that $M_{\rm orb}=m_1(\eta_1\phi_1+\eta_2\phi_2+\eta_3\phi_3)+m_2(\eta_1\phi_2\phi_3+{\rm{cyclic}})+m_3\eta_1\eta_2\eta_3$. 
Here, $\eta_a$ and $\phi_a$ represent the LC and BO order parameters in AB ($a=1$), BC ($a=2$), and CA ($a=3$) directions, respectively. The $m_1$ term, which represents the bilinear coupling between the BO and LC order, causes a large $M_{\rm{orb}}$ in the BO$+$LC coexisting state.
Therefore, the calculated $\langle L^z \rangle$ is in agreement with previous studies and supports the prediction. 

For the site-resolved quantity shown in Figs. \ref{fig:OAM} (c) and (d) with $\eta=0.01$, we define
$\langle L^z (m)\rangle=\sum_{ij}\langle c_{mi}^{\dagger}(l^z)_{ij}c_{mj}\rangle$, where $m$ labels the V site as A$_1$, B$_1$, $\cdots$, and C$_4$. Thus, $\langle L^z(m)\rangle$ represents the local orbital angular momentum induced at each V site.
This quantity may be associated with the local magnetic field generated by the loop-current pattern discussed in Ref. \cite{LC-Mag-AHE}.
Figures \ref{fig:OAM}(c) and (d) show $\langle L^z(m)\rangle$ for $\phi=0$ and $\phi=0.01$, respectively.
In both cases, the number of sites with positive and negative $\langle L^z(m) \rangle$ is equal. 
For $\phi=0$, $\vert\langle L^z(m) \rangle\vert$ shows a weak dependence on $m$, whereas for $\phi=0.01$, a slight but discernible dependence emerges. Consequently, $\langle L^z \rangle$ is considered to become larger at $\phi=0.01$.

Note that the LC order preserves translational symmetry when the $2\times2$ unit cell is used. Consequently, it is mathematically proven that no net current can be generated across a macroscopic sample (Bloch–Bohm theorem \cite{Bloch}.)

\subsection{VI. ORBITAL TRANSPORT IN LOOP-CURRENT PHASE} \vspace{-0.5em}

Finally, we consider the symmetric part $O^+_{\mu\nu}$ of the conductivity tensor.
As discussed in Sec. IV, the high-symmetry kagome phase exhibits the conventional OHE characterized by the antisymmetric transverse component $O^-_{xy}$.
The LC and BO patterns lower the magnetic and crystalline symmetries and can allow additional symmetric components of the orbital conductivity tensor.
We therefore focus here on symmetry-enabled $O^+$ responses in the LC-related phases.
Importantly, the antisymmetric part in Eq. (\ref{eqn:T-even}) has $\tau^0$ behavior, while the symmetric component in Eq. (\ref{eqn:T-odd}) is proportional to $\tau^1$ \cite{OHE-formula, OHE-tor-formula, OHE-two-terms, OHE-derive}. Therefore, the relation $|O^+_{xy}|\gg|O^-_{xy}|$ is expected to be realized in a good metal with a large $\tau$.

\begin{figure}[htp]
\includegraphics[width=.8\linewidth]{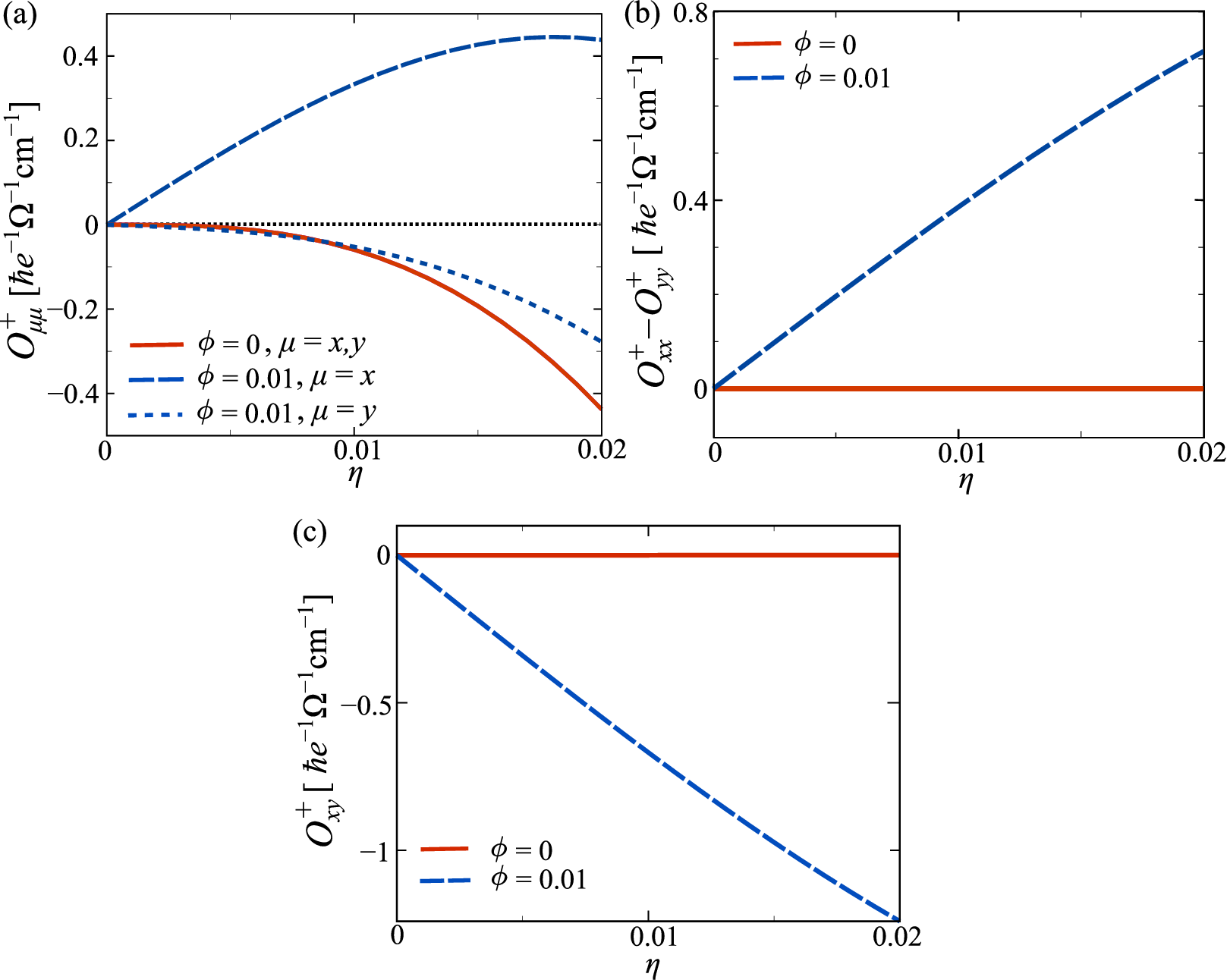}
\caption{
Symmetric components of the orbital conductivity tensor $O^+_{\mu\nu}$ as functions of the LC order parameter $\eta$ for $\gamma = 0.005$. The solid curves denote the LC-only state ($\phi = 0$), and the non-solid curves denote the BO+LC coexisting state ($\phi = 0.01$), with the dashed and dotted curves in panel (a) corresponding to $\mu = x$ and $\mu = y$, respectively. (a) Longitudinal components $O^+_{xx}$ and $O^+_{yy}$. The response is isotropic in the LC-only state and anisotropic in the BO+LC coexisting state. (b) Anisotropy of the longitudinal response, $O^+_{xx} - O^+_{yy}$. (c) Symmetric off-diagonal component $O^+_{xy}$.
}
\label{fig:OrbTran}
\end{figure}

Meanwhile, the longitudinal transport coefficient plays a significant role in applications of orbital angular momentum. In orbital torque generation, both the orbital Hall effect coefficient and the longitudinal transport determine the orbital torque strength \cite{OHE-app1,OHE-app2, OHE-OT1, OHE-OT2}.

Figure \ref{fig:OrbTran} shows the symmetric component of conductivity tensors for (a) the longitudinal transport coefficient $O^{+}_{\mu\mu}$ in the $x$ and $y$ directions, (b) the difference between the coefficient in the $x$ and $y$ directions $O^{+}_{xx}-O^{+}_{yy}$, and (c) the orbital Hall conductivity $O^{+}_{xy}$ proportional to $\tau^1$. 
We compare two cases: one with only the LC state and the other with BO$+$LC. A previous study showed that nematicity emerges in the coexistence of the BO and LC, leading to the reduction of the $C_6$ symmetry to the $C_2$ symmetry \cite{kagome-Z3}.

The absolute value of the longitudinal transport coefficient $O^{+}_{\mu\mu}$ in Fig. \ref{fig:OrbTran} (a) increases with the order parameter of the LC. 
We observe that $O^{+}_{\mu\mu}\propto\eta^3$ for $\phi=0$, whereas $O^{+}_{\mu\mu}\propto\eta$ for $\phi\ne0$. This behavior is similar to that of $M_{\rm orb}$ \cite{LC-Mag-AHE}.
Moreover, in contrast to the isotropy of the pure LC state, the BO$+$LC coexisting phase demonstrates clear anisotropy. Figure \ref{fig:OrbTran} (b) more directly illustrates the anisotropy, which progressively becomes stronger as the LC order strengthens.
In addition, the symmetric component $O^{+}_{xy}$ shown in Fig. \ref{fig:OrbTran}(c) is finite only in the nematic coexisting state and increases with the degree of anisotropy.
This behavior is consistent with symmetry. In the pure LC state, the in-plane symmetry still forbids the symmetric off-diagonal response $O^+_{xy}$, whereas the BO$+$LC coexisting state lowers the symmetry to the nematic $C_2$ state and allows a finite $O^+_{xy}$.

Although the obtained $\vert O^{+}_{xy}/\sigma^{+}_{xx}\vert$ is only on the order of $10^{-4} \sim 10^{-5} [\hbar/e]$, it could provide a possible transport signature of the LC-related ordered state in the kagome metal CsV$_3$Sb$_5$. The present study suggests that CsV$_3$Sb$_5$ may provide a useful platform for exploring symmetry-dependent orbital transport.


\subsection{VII. SUMMARY} \vspace{-0.5em}

In this paper, we investigated orbital-angular-momentum-related properties of the $135$ kagome-metal family, where kagome-derived multi-sublattice bands, vHSs, and strong $d$-$p$ hybridization provide a natural platform for orbital transport.
Without anomalous-position correction, we present a large positive OHE in CsTi$_3$Bi$_5$ and negative OHEs in CsV$_3$Sb$_5$ and CsCr$_3$Sb$_5$. 
This sign difference corresponds to a reversal of the orbital-current polarity, which is relevant to transverse $L^z$ accumulation and inverse-OHE measurements.

The orbital-sector decomposition shows that the compound and filling dependence of the OHE reflects the balance among different orbital-sector contributions.
The $\vert l^z_d \vert=2$ $d$-orbital channel gives a positive contribution, whereas the $\vert l^z \vert=1$ $d$- and $p$-orbital channels can give negative contributions.
The momentum-space orbital Berry-curvature analysis further points to the relevance of $p$-$d$-hybridized band regions.

The control calculation for CsV$_3$Sb$_5$ further shows that strong $p$-$d$ hybridization quantitatively modulates this balance, enhances the negative $p$-orbital contribution, and drives the total OHE negative in the original model.
Future work should examine the anomalous-position correction, especially its effect on the sizable $p$-orbital contribution and the OHE sign.

We also studied orbital phenomena induced by loop-current order in CsV$_3$Sb$_5$. 
The loop-current state generates finite local atomic orbital angular momentum, and the BO$+$LC coexisting state enhances the uniform component through symmetry-allowed coupling. 
In addition, the lowered symmetry in the LC-related phases allows finite symmetric components of the orbital conductivity tensor.

These results highlight kagome metals as a promising platform for orbitronics based on orbital currents and orbital angular momentum in strongly correlated systems.

\subsection{ACKNOWLEDGMENTS}
This study has been supported by Grants-in-Aid for Scientific
Research from MEXT of Japan (JP18H01175, JP20K03858, JP20K22328, JP22K14003, JP23K03299), and by the Quantum Liquid Crystal No. JP19H05825 KAKENHI on Innovative Areas from JSPS of Japan.


\subsection{Appendix A: Band structure by DFT calculation}

\begin{figure}[htp]
\includegraphics[width=.8\linewidth]{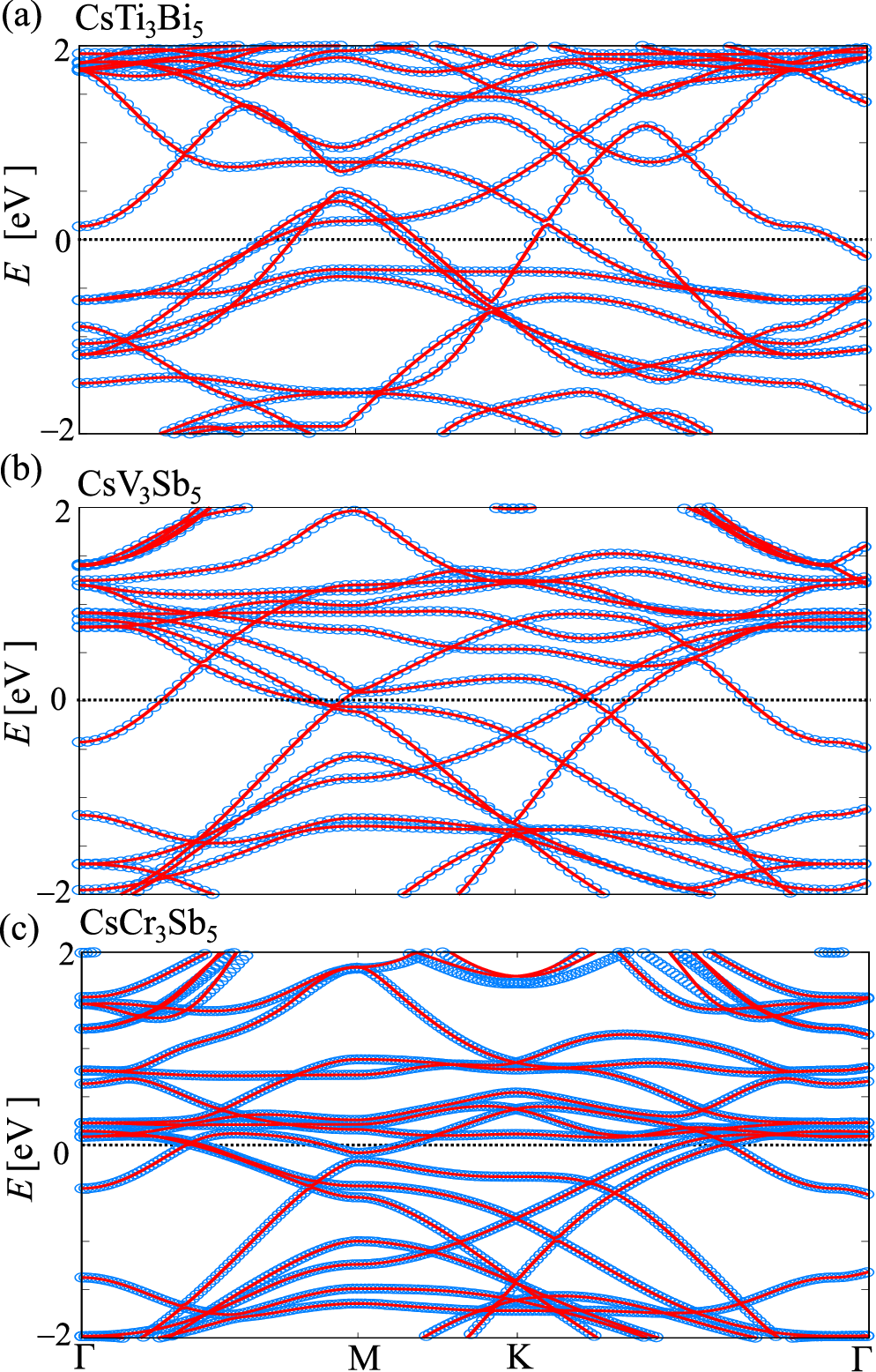}
\caption{
Comparison between the band structures obtained from first-principles calculations
(open circles) and the three-dimensional tight-binding model (solid curves)
for (a) CsTi$_3$Bi$_5$, (b) CsV$_3$Sb$_5$, and (c) CsCr$_3$Sb$_5$.
}
\label{fig:dft}
\end{figure}

Figure \ref{fig:dft} shows the comparison between the band structures obtained from first-principles calculations and those obtained from the corresponding three-dimensional tight-binding models for (a) CsTi$_3$Bi$_5$, (b) CsV$_3$Sb$_5$, and (c) CsCr$_3$Sb$_5$, respectively. The blue circles and red lines represent the first-principles and tight-binding results, respectively. The three-dimensional tight-binding models present band structures in good agreement with the first-principles results in the energy window shown. In the main-text calculations, we use the corresponding two-dimensional models by neglecting the small interlayer hopping integrals for simplicity.


\subsection{Appendix B: Conductivity formula}

Here, we explain the conductivity formula used in the main text.
Starting from the Kubo linear-response theory, the conductivity tensor in an external field $E(t)=Ee^{-i\omega t}$ is given by $\sigma_{\mu\nu}=\frac1{V}\frac1{i\omega}\Pi^{R}_{\mu\nu}(\omega)$, where the retarded correlation function is given by the Fourier transform of 
\begin{equation}
    \Pi^{R}_{\mu\nu}(t-t')=-\frac{i}{\hbar}\theta(t-t')\langle\lbrack J_{\mu}(t),J_{\nu}(t') \rbrack \rangle,
\end{equation}
which, in the eigenstate representation, is expressed as: 
\begin{align*}
    {\rm{Im}}\ \Pi^R_{\mu\nu}(\omega)=\frac{1}{4\pi\hbar}&\sum_{mn}\int_{-\infty}^{\infty}d\varepsilon \lbrack f(\varepsilon)-f(\varepsilon+\omega)\rbrack \\
    &\times J_{\mu}^{mn}\lbrack G^R(\varepsilon_m+\omega)-G^A(\varepsilon_m+\omega)\rbrack \\
    &\times J_{\nu}^{nm}\lbrack G^R(\varepsilon_n)-G^A(\varepsilon_n)\rbrack.
\end{align*}
Here, $\omega$ and $\varepsilon_{m}$ are the frequencies, $m$ and $n$ are the band indices, and $f$ is the Fermi distribution function. $J_{\mu(\nu)}$ is the current operator expressed as $J_{\mu}=-ev_{\mu}$, where $-e(e>0)$ is the electron charge and $v_{\mu}=\partial H/\partial k_{\mu}$ is the velocity. The expression is for electric current. To calculate the orbital flow, we replace $J_{\mu}$ with $J^{O^z}_{\mu}=(v_{\mu}l^z+l^z v_{\mu})/2$.
The conductivity tensor is expressed as:
\begin{align*}
    \sigma_{\mu\nu}(\omega)=&\frac1{4\pi\hbar V}\int d\varepsilon \frac{f(\varepsilon)-f(\varepsilon+\omega)}{\omega}\\
    &\times{\rm{Tr}}\lbrack J_{\mu}(G^R_{\varepsilon+\omega}-G^A_{\varepsilon+\omega})J_{\nu}(G^R_{\varepsilon}-G^A_{\varepsilon})\rbrack.
\end{align*}
In the $\omega\to 0$ limit, this expression can be separated into two parts \cite{SHE-Bi1, OHE-4d5d}:
\begin{align}
    \sigma_{\mu\nu}^I=-\frac1{V\hbar}\sum_k\int d\varepsilon \frac{df}{d\varepsilon}
    {\rm{Tr}}\lbrack J_{\mu}G^R_{\varepsilon}J_{\nu}G^A_{\varepsilon}+J_{\mu}G^A_{\varepsilon}J_{\nu}G^R_{\varepsilon}\rbrack 
\end{align}
\begin{align}
    \sigma_{\mu\nu}^{II}=&\frac1{V\hbar}\sum_k\int d\varepsilon f(\varepsilon) \notag \\
    &\times{\rm{Tr}}\lbrack J_{\mu}G^R_{\varepsilon}J_{\nu}(\partial_{\varepsilon} G^R_{\varepsilon})-J_{\mu}G^A_{\varepsilon}J_{\nu}(\partial_{\varepsilon} G^A_{\varepsilon})-\langle \mu\leftrightarrow\nu\rangle \rbrack
\end{align}
where $I$ and $II$ represent the Fermi-surface term and the Fermi-sea term introduced in the main text.

By following Refs. \cite{OHE-formula, OHE-tor-formula, OHE-two-terms}, we obtain:
\begin{align}
    &\sigma^{+}_{\mu\nu}=-\frac{\hbar}{\pi V}\sum_{knm}
          \frac{\gamma^2 {\rm{Re}}\lbrack \langle n\vert J_{\mu} \vert m \rangle \langle m\vert J_{\nu} \vert n \rangle\rbrack}{(E_{kn}^2+\gamma^2)(E_{km}^2+\gamma^2)}.
\end{align}
\begin{align}
    \sigma^{-}_{\mu\nu}&=-\frac{\hbar}{2\pi V}\sum_{k,n\ne m}
    {\rm{Im}}\lbrack \langle n\vert J_{\mu} \vert m \rangle \langle n\vert J_{\nu} \vert m \rangle\rbrack \notag \\
        &\times\lbrace\frac{\gamma(E_{km}-E_{kn}) }{(E_{kn}^2+\gamma^2)(E_{km}^2+\gamma^2)} \notag \\
        &+\frac{2\gamma}{(E_{kn}-E_{km})(E_{km}^2+\gamma^2)} \notag \\
        &+\frac{2\gamma}{(E_{kn}-E_{km})^2} {\rm Im} \left[{\rm ln}\frac{E_{km}-i\gamma}{E_{kn}-i\gamma}\right]
        \rbrace.
\end{align}

In the limit $\gamma\to 0$, the two components become \cite{OHE-formula}:
\begin{align*}
    \sigma^{+}_{\mu\nu}=-&\frac{\hbar}{2\gamma V}\sum_{kn} \langle n\vert J_{\mu} \vert n \rangle \langle n\vert J_{\nu} \vert n \rangle \delta(E_{kn}) \\
    \sigma^{-}_{\mu\nu}=-&\frac{2\hbar}{V}\sum_k\sum_{n}^{\rm{occ}}\sum_{m\ne n} {\rm{Im}}\left[ \frac{\langle n\vert J_{\mu} \vert m \rangle \langle m\vert J_{\nu} \vert n \rangle}{(E_{kn}-E_{km})^2} \right].    
\end{align*}
which are the expressions in Eqs. (\ref{eqn:T-odd}) and (\ref{eqn:T-even}).
Meanwhile, we can easily see the $1/\gamma$ behavior of the symmetric component $\sigma^{+}_{\mu\nu}$ in the limit $\gamma \to 0$, which indicates a giant $\sigma^{+}_{\mu\nu}$ in a metal with a large relaxation time $\tau\equiv \hbar/2\gamma$.

\subsection{Appendix C: Damping-Rate Dependence of the Orbital Hall Effect}

\begin{figure}[htp]
\includegraphics[width=.8\linewidth]{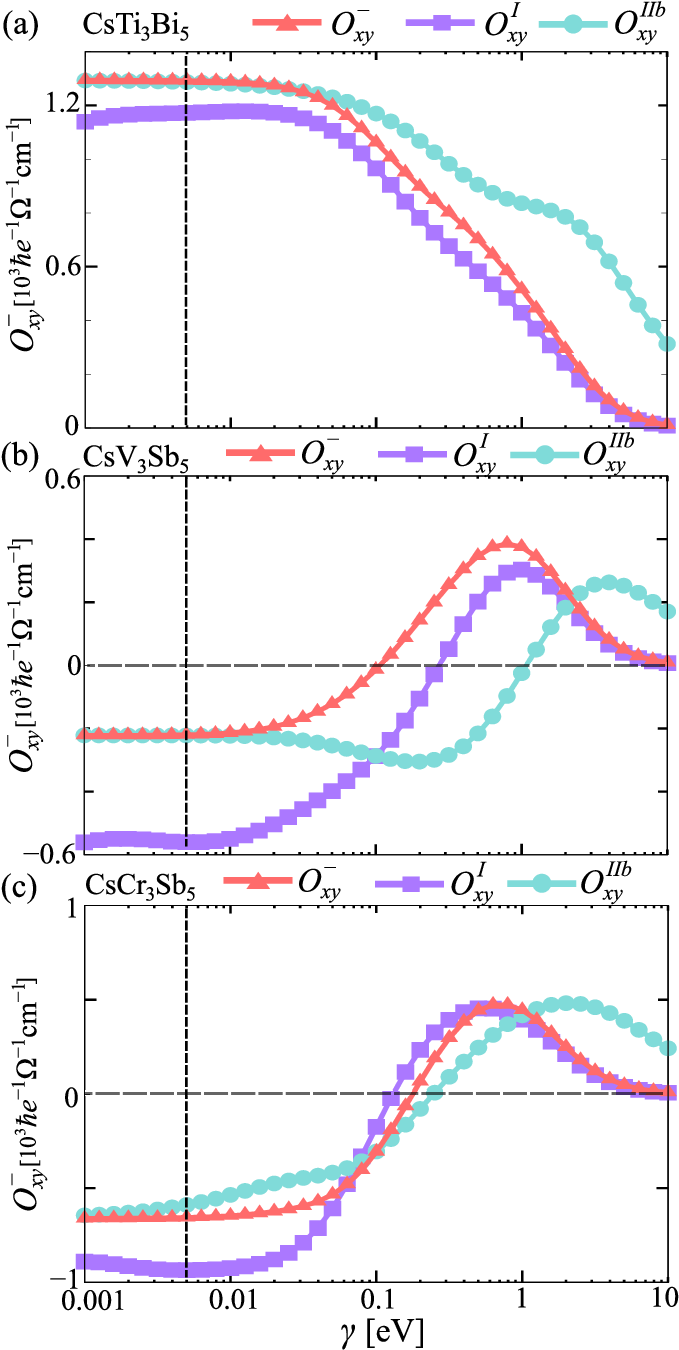}
\caption{
$\gamma$ dependence of orbital Hall effect $O^{-}_{xy}$, Fermi surface term $O^I_{xy}$, and Fermi sea term $O^{IIb}_{xy}$ in (a) CsTi$_3$Bi$_5$, (b) CsV$_3$Sb$_5$, and (c) CsCr$_3$Sb$_5$ models. The three quantities are distinguished by the symbols indicated in the legend. The dashed line indicates the value $\gamma=0.005$ used in the calculation. In the low-damping-rate region the relation $O^-_{xy}=O^{IIb}_{xy}$ holds well, while $O^-_{xy}=O^I_{xy}$ is valid for large $\gamma$.
}
\label{fig:g-dep}
\end{figure}

We explained the relation between $O^{-}_{xy}$, $O^I_{xy}$, and $O^{IIb}_{xy}$ in the formula section. Here, we show their damping-rate dependence more clearly and verify that explanation.

Figure \ref{fig:g-dep} exhibits the antisymmetric orbital Hall effect $O^{-}_{xy}$, the Fermi surface term $O^I_{xy}$, and the Fermi sea term $O^{IIb}_{xy}$ as functions of the quasiparticle damping rate $\gamma$. We use the value $\gamma=0.005$ for the calculation, which is represented by the dashed line. 
At this value, the relation $O_{xy}\approx O^{IIb}_{xy}$ holds well, validating the discussion of the Fermi sea term $O^{IIb}_{xy}$ in the main text. 
In contrast, the relation $O_{xy}\approx O^I_{xy}$ is valid in the high damping rate region ($\gamma>0.5$), which is consistent with previous studies.



\subsection{Appendix D: Anomalous position effect}

Recent studies have shown that the anomalous position correction can substantially affect the OHE \cite{APE-d, APE-Si}.
In the main text, we evaluate the orbital-current operator without the anomalous-position correction.
We briefly outline this correction here to clarify the scope of this approximation.

The anomalous position is a correction to the canonical position operator arising from dipole matrix elements between Wannier basis functions \cite{APE-d}.
Because this correction is orbital dependent, it can modify the velocity and orbital-current operators and thereby change the magnitude, and in some cases even the sign, of the OHE.
This effect is expected to be particularly important for spatially extended $p$ orbitals \cite{APE-Si}.

We next summarize its formal expression in the Wannier basis.
Let $\{|l\R\rangle\}$ be a set of Wannier states, where $l$ is the Wannier index, including the orbital and sublattice degrees of freedom, and $\R$ is a Bravais lattice vector. 
In this basis, the position operator can be decomposed as
\begin{equation}
    \hat{\r}=\hat{\r}_0+\delta\hat{\r},
\end{equation}
where
\begin{equation}
    \hat{\r}_0=\sum_{l\R}|l\R\rangle \R \langle l\R|
\end{equation}
describes the position of the unit-cell center, while
\begin{equation}
    \delta\hat{\r}
    =\sum_{l\R,l'\R'} |l\R\rangle
    \delta r_{ll'}(\R'-\R)
    \langle l'\R'|
\end{equation}
describes the displacement from the unit-cell center. 
The dipole matrix element is defined as
\begin{equation}
    \delta r_{ll'}(\R'-\R)
    =
    \langle l\R|(\hat{\r}-\R)|l'\R'\rangle
    =
    \langle l{\bf 0}|\hat{\r}|l'(\R'-\R)\rangle .
\end{equation}
Thus, $\delta\hat{\r}$ contains information on the spatial distribution of the Wannier basis functions, which is absent in a simple tight-binding treatment.

Introducing the Bloch sum of Wannier states,
\begin{equation}
    |l\k\rangle=\frac{1}{\sqrt{N}}\sum_{\R}e^{i\k\cdot\R}|l\R\rangle,
\end{equation}
the position operator in the Wannier representation momentum space is written as
\begin{equation}
    \hat{\r}_{ll'}(\k)=i\delta_{ll'}\nabla_{\k}+A_{ll'}(\k),
\end{equation}
where
\begin{equation}
    A_{ll'}(\k)=\sum_{\R}e^{i\k\cdot\R}
    \langle l{\bf 0}|\hat{\r}|l'\R\rangle
\end{equation}
is the anomalous position. This expression is equivalent to the definition of the Berry connection under the periodic Bloch-function basis $\vert u_{l\k}\rangle=e^{-i\k\cdot{\bf r}}|l\k\rangle$:
\begin{equation}
    A_{ll'}(\k)=i\langle u_{l\k}\vert \nabla_{\k}u_{l'\k}\rangle.
\end{equation}

Therefore, the velocity operator becomes
\begin{equation}
    \hat{\v}(\k)
    =
    \frac{1}{i\hbar}[\hat{\r}(\k),\hat H(\k)]
    =
    \frac{1}{\hbar}\nabla_{\k}\hat H(\k)
    +\frac{1}{i\hbar}[A(\k),\hat H(\k)] .
\end{equation}
The first term is the conventional group-velocity-like term used in the main-text calculation, whereas the second term is the anomalous-position contribution to the velocity. 
In the present work, this second term has been ignored. 
A calculation including this contribution is left for future work.


\subsection{Appendix E: Bond order and Loop current order}

In the main text, we introduce bond order (BO) and loop current order (LC) by real hopping $\delta t^b=\pm \phi$ and imaginary hopping $\delta t^c=\pm i\eta$. We explain them more detail here.

\begin{figure}[htp]
\includegraphics[width=.8\linewidth]{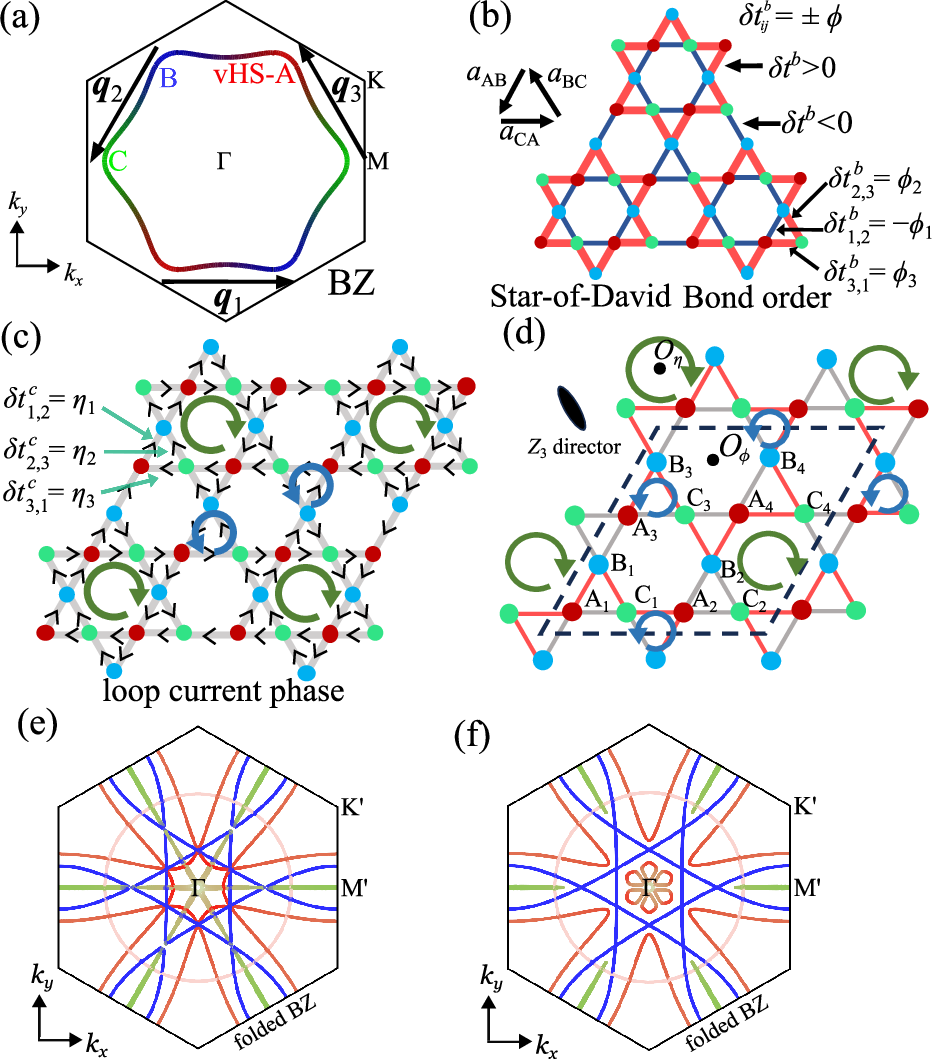}
\caption{
(a) $d_{xz}$-orbital Fermi surface with inter-vHS nesting vectors $\q_n$ ($n=1$--$3$). The sublattice components of the Fermi surface are indicated in the panel. 
(b) Triple-$\q$ Star-of-David bond order. The two bond modulations, $\delta t^b>0$ and $\delta t^b<0$, are indicated in the panel. 
(c) Triple-$\q$ loop-current (LC) order induced by the imaginary hopping modulation. Clockwise and anticlockwise LCs are indicated by the directions of the arrows. 
(d) $2\times2$ unit-cell kagome lattice model with the BO+LC phase on the V sites. The directions of the arrows indicate the clockwise and anticlockwise orbital loop currents. 
(e) Folded Fermi surfaces without BO or LC order ($\phi=\eta=0$). 
(f) Fermi surfaces with $\phi=\eta=0.01\,\mathrm{eV}$.
}
\label{fig:order}
\end{figure}

Theoretical works have successfully explained the BO or LC state by applying the "paramagnon interference mechanism", which reveals that the hopping integral modulation is the symmetry breaking in the self-energy and is derived from the density-wave (DW) equation \cite{Tazai-DW, kagome-Z3}. This theoretical approach has also been applied effectively to Fe-based and cuprate superconductors and twisted bilayer graphene \cite{FeSe,TBG,Cu-SC}.

The wavevectors of BO and LC order correspond to the inter-sublattice nesting vectors $\q_n$ ($n=1-3$) in Fig. \ref{fig:order} (a). Here we only present the $d_{xz}$ Fermi surface, which lies on the three vHS points and is composed of a single sublattice A, B, or C. 
The triple-$\q$ BO of the Star-of-David pattern is shown in Fig. \ref{fig:order} (b), with red (blue) denoting the slight accumulation (depletion) of charge. Figure \ref{fig:order} (c) demonstrates the triple-$\q$ LC order, which holds the odd parity relation $\delta t^c_{ij}=-\delta t^c_{ji}$. 
We can define the order parameter set ${\bm \phi}\equiv(\phi_1,\phi_2,\phi_3)$ and ${\bm \eta}\equiv(\eta_1,\eta_2,\eta_3)$ with the wavevector $\q_n$, as illustrated in Figs. \ref{fig:order} (b) and (c). With the same strength modulation applied to all $\q_n$, the order parameters in Figs. \ref{fig:order} (b) and (c) can be written as $(\phi,\phi,-\phi)$ and $(\eta,\eta,\eta)$. 

The LC order together with the Star-of-David BO leads to the nematic coexisting state in Fig. \ref{fig:order} (d). The unit cell in the BO$+$LC state is magnified by $2\times2$ times. Figure \ref{fig:order} (e) exhibits the folded Fermi surfaces of the $12$-site lattice model for $\phi=\eta=0$ with numerous band crossings. Upon applying an order with $\phi=\eta=0.01$, the Fermi surface is reconstructed as in Fig. \ref{fig:order} (f), indicating a redistribution of the Berry curvature.




\end{document}